# *Gain-Momentum Locking in Chiral-Gain Media*

*João C. Serra[1,]*, Nader Engheta[2], Mário G. Silveirinha[1,]†*

[(1)] *University of Lisbon–Instituto Superior Técnico and Instituto de Telecomunicações, Avenida Rovisco Pais, 1, 1049-001 Lisboa, Portugal*
[(2)] *University of Pennsylvania, Department of Electrical and Systems Engineering, Philadelphia, PA 19104, U.S.A.*

**Abstract**

Conventional optical materials are characterized by either a dissipative response, which results in polarization-independent absorption, or by a gain response that leads to wave amplification. In this work, we study a peculiar class of materials with chiral-gain properties, where gain selectively amplifies waves of one polarization handedness, while dissipation suppresses the opposite handedness. We uncover a novel phenomenon, "gain-momentum locking", at the boundary of chiral-gain media, where surface plasmons are amplified or attenuated based on their direction of propagation. This effect, driven by the interplay between spin-momentum locking and polarization-sensitive non-Hermitian responses, enables precise control over unidirectional wave propagation. Our findings open the door to photonic devices with unprecedented capabilities, such as lossless unidirectional edge-wave propagation and the generation of light with intrinsic orbital angular momentum.

---

* E-mail: joao.serra@lx.it.pt
† To whom correspondence should be addressed: mario.silveirinha@tecnico.ulisboa.pt

# Main Text

The propagation properties in a non-magnetic medium are tailored by the photonic dispersion and isofrequency surfaces, governed by the Hermitian part $\overline{\varepsilon}' = \left(\overline{\varepsilon} + \overline{\varepsilon}^{\dagger}\right)/2$ of the permittivity tensor $\overline{\varepsilon}$. When all eigenvalues of $\overline{\varepsilon}'$ are positive, the isofrequency contours assume an ellipsoidal shape, indicative of dielectric-type behavior. Conversely, if all eigenvalues are negative, the material behaves akin to a metal, not supporting wave propagation. More uniquely, when the tensor possesses both positive and negative eigenvalues, the medium displays a dual response characterized by direction-dependent band gaps. These are known as hyperbolic-type materials, such as hexagonal boron nitride. Within this category, type I and type II systems have one or two negative eigenvalues, respectively [1-3].

Evidently, the response of realistic materials is usually non-Hermitian. In such cases, the material response can be conveniently decomposed as $\overline{\varepsilon} = \overline{\varepsilon}' + i\overline{\varepsilon}''$, where the Hermitian part of the response $\overline{\varepsilon}'$ is defined as before and the non-Hermitian part is given by $\overline{\varepsilon}'' = \left(\overline{\varepsilon} - \overline{\varepsilon}^{\dagger}\right)/\left(2i\right)$. Both components ($\overline{\varepsilon}'$ and $\overline{\varepsilon}''$) are Hermitian tensors with real-valued eigenvalues. The non-Hermitian component $\overline{\varepsilon}''$ governs the power transfer between the medium and the wave. Specifically, in the time-harmonic regime, the average power per unit volume transferred from the wave to the material (dissipated power) is given by $p_{\mathrm{d}} = \frac{1}{2}\omega\varepsilon_0 \mathbf{E}^* \cdot \overline{\varepsilon}'' \cdot \mathbf{E}$ [4]. This formula holds true for dispersive and nonlocal systems. For passive materials, the eigenvalues must be positive to ensure $p_{\mathrm{d}} > 0$. Conversely, in standard gain materials (e.g., within a laser cavity), when gain overcomes dissipative effects, $p_{\mathrm{d}} < 0$ and all the eigenvalues are negative.

Here, we consider an intermediate scenario where $\bar{\varepsilon}''$ is indefinite, having eigenvalues of both signs. Under such conditions, the material can exhibit either gain or loss depending on the wave polarization. This possibility was recently explored in the context of a photonic analogue of semiconductor transistors [5]. Additionally, it has been suggested that indefinite-gain responses can be implemented in electrically biased low-symmetry materials with nontrivial Berry curvature dipoles [6, 7], as well as in related optically pumped platforms [8].

In electrical realizations, the required gain is drawn from a drift current via a nonlinear process that couples the Bloch velocity to the applied electric field through the electronic Berry curvature [6] (anomalous velocity term). When an external bias drives carriers through a crystal lacking inversion symmetry, the anomalous velocity yields polarization-dependent amplification or dissipation. Symmetry analysis shows that materials from different point groups can host such chiral-gain [9], with trigonal tellurium being the most well-studied example [7]. Moreover, engineered systems like twisted bilayer graphene, which exhibit large Berry curvature dipoles, are promising platforms for its experimental realization [6].

Another route to chiral-gain employs optical pumping of low-dimensional systems [8]. Transition-metal dichalcogenide monolayers feature two inequivalent Dirac valleys (K and K') whose Bloch states carry opposite intrinsic chirality. Circularly polarized pump light preferentially excites one valley, and under strong inversion, it produces a population imbalance between valleys [10-12]. The resulting non-Hermitian response may be polarization-sensitive, leading to selective amplification of one circular polarization over the other [8]. Notably, although non-chiral, optical gain has already been demonstrated in these systems [13].

Finally, we note in passing that the chiral-gain concept may extend far beyond photonics. For example, in mechanically active metamaterials, the concept of "odd elasticity" introduced by Vitelli and collaborators can be viewed as chiral-gain for elastic waves. In such systems, actuators or motors supply directional energy transfer that locks gain to "polarization" [14, 15].

The main objective of this Letter is to highlight a universal effect of "gain-momentum" locking that naturally emerges at the interface of a truncated chiral-gain medium, regardless of the specific wave system or physical implementation. This effect arises from the intrinsic interplay between spin, momentum, and gain in such media. As this interplay is grounded in universal principles of wave physics, gain–momentum locking offers a unifying framework for understanding non-Hermitian effects across optics, acoustics, mechanics, and beyond.

To develop these ideas, we consider, without loss of generality, an active optical material described by the following permittivity:

$$\bar{\varepsilon} = \begin{pmatrix} \varepsilon_{xx} & 0 & \varepsilon_{xz} \\ 0 & \varepsilon_{yy} & 0 \\ 0 & 0 & \varepsilon_{zz} \end{pmatrix}. \tag{1}$$

As outlined in the Introduction, and as will be discussed in more detail later, related responses can be engineered either through electrical or optical pumping [6-8]. First, we assume that the permittivity tensor (1) is real-valued and frequency-independent. In such a case, the non-Hermitian component $\bar{\varepsilon}''$ is anti-symmetric and it can be written in terms of the vector $\mathbf{\Omega} = +\frac{\varepsilon_{xz}}{2}\hat{\mathbf{y}}$ as $\bar{\varepsilon}'' = -i\mathbf{\Omega}\times\mathbf{1}$. We shall refer to $\mathbf{\Omega}$ as the *gain vector*. The nontrivial eigenvalues of $\bar{\varepsilon}''$ are $\pm\frac{\varepsilon_{xz}}{2}$ and the corresponding eigenvectors are the circular polarization states $\mathbf{e}_{\pm} = \frac{1}{\sqrt{2}}\left(\hat{\mathbf{x}} \pm i\hat{\mathbf{z}}\right)$.

The non-Hermitian response in this system has chiral properties as the power transfer is governed by the quadratic form $p_{\mathrm{d}} = \frac{-1}{2}\omega\varepsilon_0\left|\mathbf{E}\right|^2\mathbf{\Omega}\cdot\boldsymbol{\sigma}$, which depends on the relative orientation of the spin angular momentum of light $\boldsymbol{\sigma} = i\left(\mathbf{E}\times\mathbf{E}^*\right)/\left|\mathbf{E}\right|^2$ with respect to the gain vector $\mathbf{\Omega}$ [16]. Remarkably, when the spin angular momentum is parallel (anti-parallel) to the gain vector, the medium supplies energy to (absorbs energy from) the wave. Therefore, when $\varepsilon_{xz} > 0$, the $\mathbf{e}_+$ polarization state activates "dissipation", whereas the orthogonal $\mathbf{e}_-$ polarization state activates optical gain.

For simplicity, we restrict our attention to waves propagating in the $xoz$ plane with the magnetic field oriented along the $y$ direction (aligned with $\mathbf{\Omega}$), so that $\mathbf{H} = H_y\left(x,z\right)\hat{\mathbf{y}}$. It is shown in the supplementary information [17] that, provided $\left|\varepsilon_{xz}\right| \le 2\sqrt{\varepsilon_{xx}\varepsilon_{zz}}$, the bulk material response is stable. In these conditions, the bulk modes in the chiral gain medium are linearly polarized and thereby are unaffected by gain or dissipation, as $p_{\mathrm{d}}$ vanishes for such a case.

Surface plasmon polaritons (SPPs) – which naturally occur at dielectric-metal interfaces – possess inherent chiral properties. The "spin" of surface plasmons is linked to their direction of propagation, through a property known as "spin-momentum locking" [18, 19]. Therefore, intuitively one may expect that materials exhibiting chiral-gain can amplify plasmons propagating along one direction whose chirality aligns with the gain vector, while suppressing counter-propagating waves with opposite chirality (Fig. 1a).

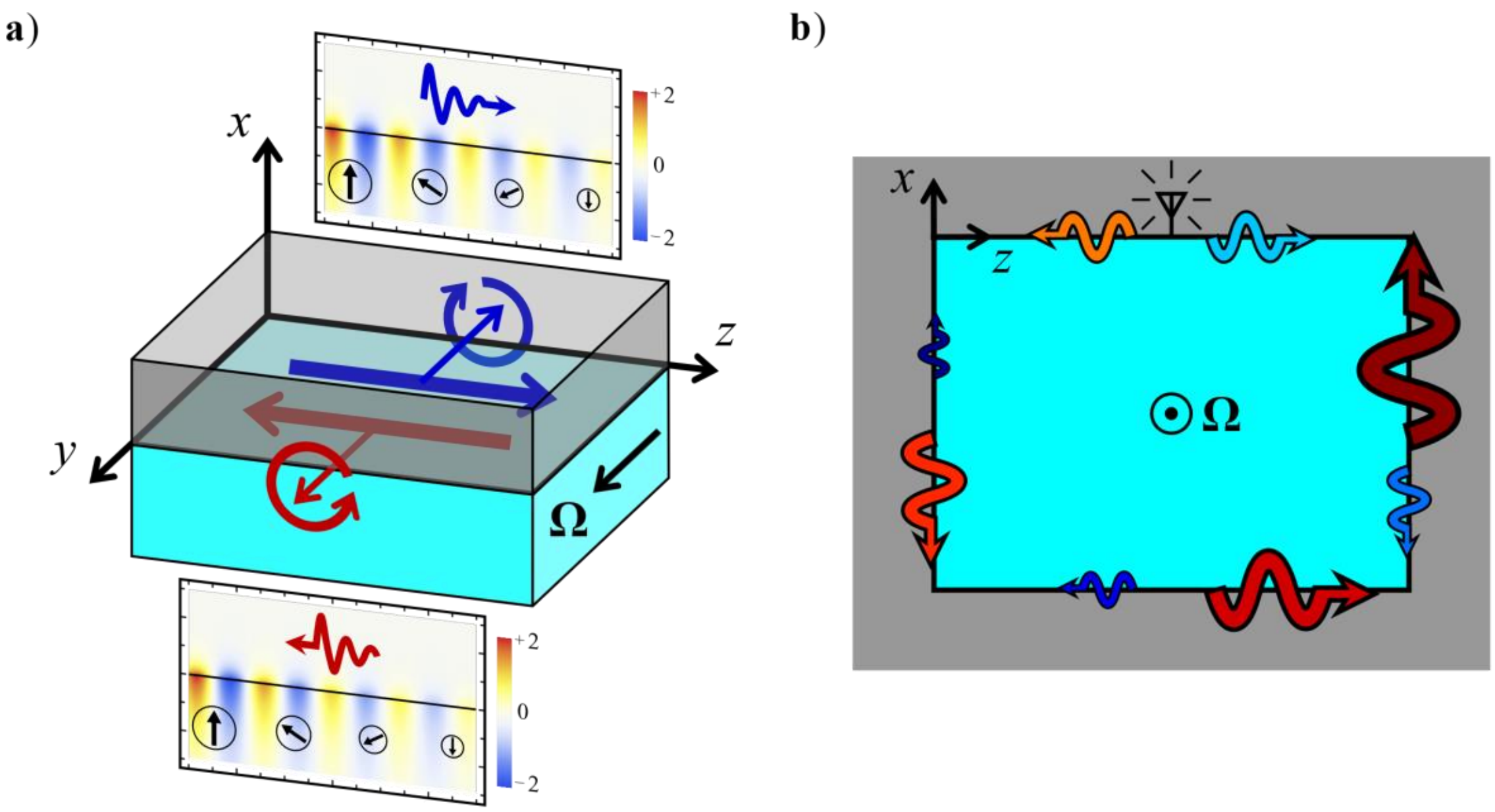

**Fig. 1 (a)** Interface between a chiral-gain medium (region $x<0$, shaded in cyan) and an isotropic material (metal; $x>0$, shaded in gray). Due to the spin-momentum locking, surface plasmons have an inherent chirality. The oriented blue and red circles represent the spin $\boldsymbol{\sigma}$ of the plasmons in the chiral-gain material side of the interface for waves propagating along $+z$ and $-z$, respectively. The chiral-gain medium amplifies plasmons whose spin aligns with the gain vector $\boldsymbol{\Omega}$ (directed along $+y$) and suppresses those with opposite spin. **(b)** Illustration of an "edge-cavity" laser based on a chiral-gain medium and a metallic coating. The wave that propagates in the counter-clockwise (clockwise) direction is amplified (attenuated). Hence, the orbital angular momentum of the lasing mode is locked to the direction of the gain vector $\boldsymbol{\Omega}$. Lighter/darker colors represent the edge states at earlier/later time instants.

In order to demonstrate such "gain-momentum locking", we consider the flat interface represented in Fig. 1a between a chiral-gain medium ($x<0$) described by the permittivity tensor (1) and a metal ($x>0$) with relative permittivity $\varepsilon_{\mathrm{m}}(\omega)=1-\dfrac{\omega_{\mathrm{m}}^{2}}{\omega(\omega+i\Gamma)}$. Here, $\omega_{\mathrm{m}}$ is the metal plasma frequency and $\Gamma$ is the collision frequency.

We start by analyzing the system in the quasi-static regime where retardation effects are negligible and the speed of light can be taken as $c=\infty$. In this regime, the magnetic field is negligible ($\mathbf{H}\approx\mathbf{0}$) and the electric field is determined by a scalar potential ($\mathbf{E}\approx-\nabla\phi$). Since the system has continuous translational symmetry in the $yoz$ plane, the electric potential associated with a surface plasmon is of the form:

$$\phi(x,z,t)=\phi_0 e^{-i\omega t}e^{iqz}\times\begin{cases}e^{+\gamma_{\mathrm{C}0}x}, & x<0\\ e^{-\gamma_{\mathrm{m}0}x}, & x>0\end{cases}, \tag{2}$$

where $q$ is the propagation constant along $z$. The electric potential must satisfy Gauss's law $\nabla\cdot\left(\overline{\varepsilon}\cdot\nabla\phi\right)=0$, which for plane waves (spatial variation $e^{i\mathbf{k}\cdot\mathbf{r}}$) reduces to $\mathbf{k}\cdot\overline{\varepsilon}\cdot\mathbf{k}=0$. From here, it is found that the attenuation constants along $z$ are $\gamma_{\mathrm{m}0}=\sqrt{q^2}$ in the metal region, and $\gamma_{\mathrm{C}0}=\sqrt{q^2\left(\dfrac{\varepsilon_{zz}}{\varepsilon_{xx}}-\dfrac{\varepsilon_{xz}^2}{4\varepsilon_{xx}^2}\right)}-iq\dfrac{\varepsilon_{xz}}{2\varepsilon_{xx}}$ in the chiral-gain region.

Let us suppose that the chiral-gain response is a weak perturbation of the response of an isotropic dielectric ($\varepsilon_{xx}\approx\varepsilon_{zz}>>|\varepsilon_{xz}|$), so that $\gamma_{\mathrm{C}0}\approx\gamma_{\mathrm{m}0}=\sqrt{q^2}$. Then, the electric field in the chiral-gain region is roughly $\mathbf{E}\approx-|q|\phi_0 e^{-i\omega t}e^{iqz}e^{+|q|x}\left[\hat{\mathbf{x}}+i\,\mathrm{sgn}(q)\hat{\mathbf{z}}\right]$, for $q$ real-valued. As expected, the orientation of the transverse spin of the plasmons, $\boldsymbol{\sigma} = -\mathrm{sgn}(q)\hat{\mathbf{y}}$ depends on the direction of propagation [18, 19]. If the gain vector $\boldsymbol{\Omega} = +\frac{\varepsilon_{xz}}{2}\hat{\mathbf{y}}$ points along $+y$, the transverse spin $\boldsymbol{\sigma}$ is parallel (anti-parallel) to $\boldsymbol{\Omega}$ for propagation along $-z$ ($+z$). Consequently, these directions of propagation correspond to conditions of plasmon amplification and attenuation, respectively, as illustrated in Fig. 1a.

To confirm this result, next we derive the quasi-static dispersion of the surface plasmons. To this end, we enforce the continuity of the normal component of the electric displacement vector at the interface ($\hat{\mathbf{x}} \cdot \mathbf{D}|_{x=0^-} = \hat{\mathbf{x}} \cdot \mathbf{D}|_{x=0^+}$ with $\mathbf{D} = -\varepsilon_0 \bar{\varepsilon} \cdot \nabla \phi$). This yields:

$$\gamma_{\mathrm{m0}} \varepsilon_{\mathrm{m}}(\omega) + \gamma_{\mathrm{C0}} \varepsilon_{xx} + iq\varepsilon_{xz} = 0. \tag{3}$$

The attenuation coefficients must reside in the complex right half-plane, i.e., $\mathrm{Re}\{\gamma_{\mathrm{m0/C0}}\} > 0$ to confine the energy at the interface $x = 0$.

Let us first analyze excitations with $q$ real-valued and $\omega = \omega' + i\omega''$. These correspond to the natural modes of a ring-type edge-resonator in the $x = 0$ plane (see [20] for a related discussion). For passive resonators, it is necessary that $\omega'' < 0$, indicating that the excitation relaxes over time ($e^{-i\omega t} = e^{-i\omega' t} e^{\omega'' t}$); conversely, $\omega'' > 0$ is indicative of a lasing mode. Solving Eq. (3) with respect to the oscillation frequency, it is found that in the $\Gamma = 0^+$ limit:

$$\omega\big|_{\mathrm{QS}} = \frac{\omega_{\mathrm{m}}}{\left(1+\sqrt{\varepsilon_{xx}\varepsilon_{zz}-\frac{\varepsilon_{xz}^2}{4}+i\,\mathrm{sgn}(q)\frac{\varepsilon_{xz}}{2}}\right)^{1/2}}. \tag{4}$$

When the chiral-gain response is nearly isotropic ($\varepsilon_{xx} \approx \varepsilon_{zz} >> |\varepsilon_{xz}|$), it follows that $\omega\big|_{\mathrm{QS}} = \omega' + i\omega'' \approx \frac{\omega_{\mathrm{m}}}{\sqrt{1+\varepsilon_{xx}}}\left(1-i\,\mathrm{sgn}(q)\frac{\varepsilon_{xz}}{4(1+\varepsilon_{xx})}\right)$. Thus, the gain rate $\omega''$ of the surface plasmon excitations is locked to the direction of propagation. Specifically, in accordance with the relative orientation between $\boldsymbol{\sigma}$ and $\boldsymbol{\Omega}$, it is observed that when $\varepsilon_{xz} > 0$, plasmons propagating along $+z$ are attenuated ($\omega'' < 0$), whereas those propagating along $-z$ are amplified ($\omega'' > 0$). Remarkably, even though each material individually exhibits a stable bulk response, their combination generates an edge excitation that grows exponentially over time, corresponding to a lasing mode.

Figure 2a displays the exact complex frequency spectrum (blue lines) calculated by solving the full Maxwell's equations and incorporating retardation effects [17, Eq. (S6)]. The black lines represent the quasi-static solution discussed before, which, as expected, accurately describes the asymptotic behavior of the plasmons for large wave numbers. Consistent with the quasi-static analysis, the sign of the gain rate $\omega''$ computed with the exact theory is locked to the sign of the propagation constant $q$. The results shown in Fig. 2a use a large gain parameter ($\varepsilon_{xz} = 0.8$) to highlight the impact of the chiral-gain.

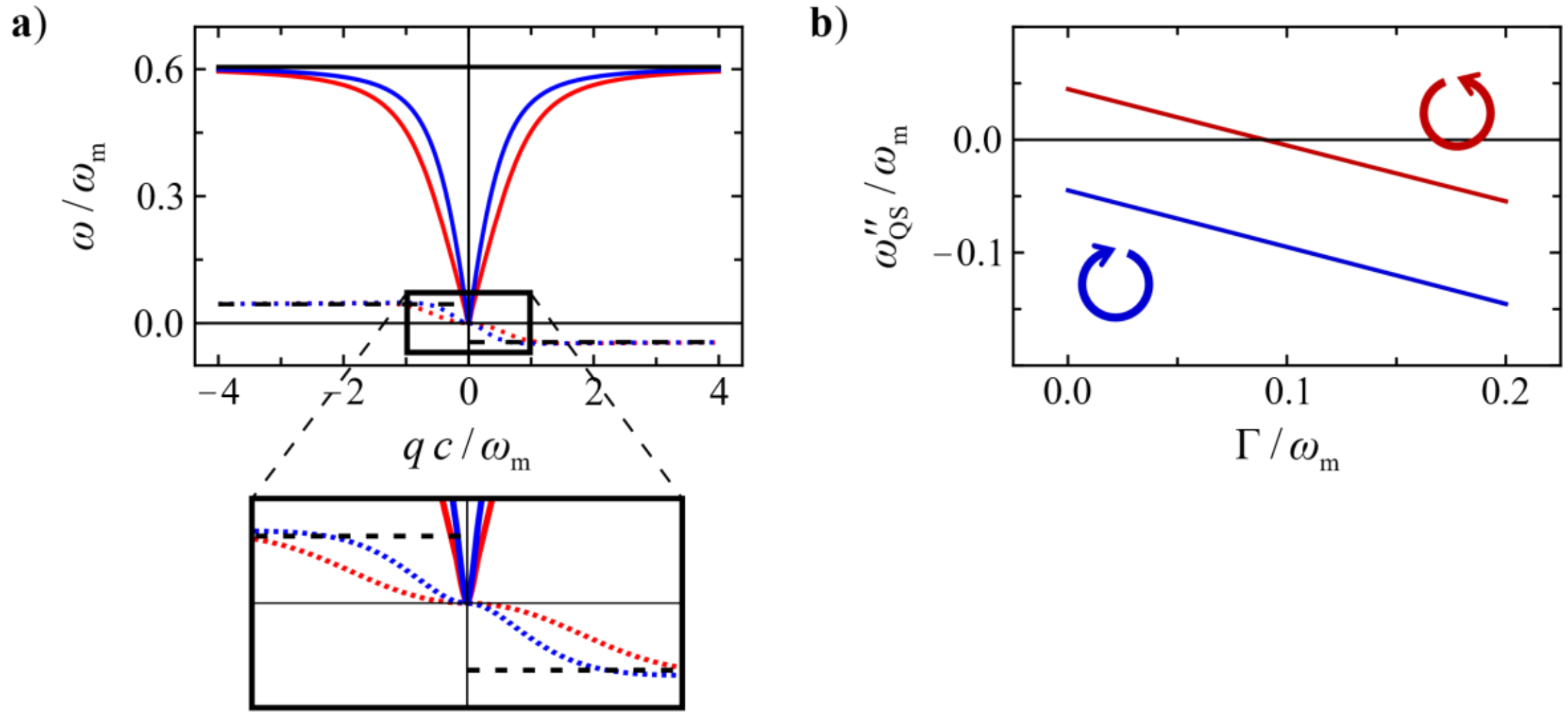

**Fig. 2 (a)** Exact dispersion of the surface plasmons supported at an interface between a chiral-gain medium and a lossless Drude-type metal ( $\Gamma = 0$ ) at the $x = 0$ (blue) and $z = 0$ (red) interfaces. The black horizontal line represents the quasi-static solution. The figure shows the real (solid) and imaginary (dashed) parts of the frequency $\omega' + i\omega''$ as a function of the wave number $q \in \mathbb{R}$ for $\varepsilon_{xx} = 1$, $\varepsilon_{zz} = 3$, $\varepsilon_{xz} = 0.8$. **(b)** Growth rate $\omega''_{\rm QS}$ for highly-confined plasmons propagating in the CCW direction with $\boldsymbol{\sigma} \sim +\boldsymbol{\Omega}$ (red) and in the CW direction with $\boldsymbol{\sigma} \sim -\boldsymbol{\Omega}$ (blue) as a function of the Drude metal loss parameter $\Gamma$. The chiral-gain material has the same parameters as in (a).

An interesting configuration for implementing an oscillator driven by the chiral-gain is illustrated in Fig. 1b. It consists of an edge-type resonator with four interfaces: two interfaces perpendicular to the $x$ axis and another two perpendicular to the $z$ axis. The propagation along the (horizontal) interfaces perpendicular to the $x$ axis has been analyzed previously. We demonstrate in the supplementary information [17] that the quasi-static dispersion of the edge states for vertical interfaces is still given by Eq. (4), with $q$ understood as the propagation constant of the edge guide, with positive values indicating clockwise propagation. Thus, the propagation along the lateral interfaces is also governed by the gain-momentum locking discussed previously.

As depicted in Fig. 1b, a field source near the top interface may excite surface plasmons propagating in both clockwise (CW) and counterclockwise (CCW) directions with respect to the gain vector $\mathbf{\Omega}$ (oriented along the $+y$ axis): while the former waves suffer attenuation, the latter are continuously amplified. Over time, the edge-resonator behaves as a unidirectional guide, because the CW modes are effectively suppressed. In practice, nonlinear effects will curb the exponential growth, resulting in a stable oscillator. Remarkably, the orbital angular momentum of the lasing modes of the edge-cavity is aligned with the direction of the gain vector $\mathbf{\Omega}$. We show in Fig. 2b that the amplification is suppressed for sufficiently high losses in the metal region.

Until now, our discussion has primarily focused on the natural modes of closed systems, such as resonators, where the boundary forms a closed loop. To investigate potential applications of chiral-gain materials in open systems, next we consider plasmons characterized by a real-valued frequency $\omega$, driven by some external time-harmonic excitation. In this context, the wave number of the plasmons becomes complex-valued: $q = q' + iq''$. We consider the same setup as in Fig. 1a, so that the propagation direction is along $z$, with a propagation factor $e^{iqz} = e^{iq'z} e^{-q''z}$. For simplicity, the metal region is assumed lossless.

Figure 3a presents a snapshot of the magnetic field distribution associated with a surface plasmon at a specific oscillation frequency ($\omega = 0.4\omega_{\mathrm{m}}$). Interestingly, the density plot is identical regardless of the propagation direction of the SPP. In the end matter, we demonstrate that this property arises from the fact that the dispersionless chiral-gain medium exhibits time-reversal symmetry. In particular, $q'' = \mathrm{Im}\{q\}$ is independent of the propagation direction. The arrows in Fig. 3a qualitatively indicate the orientation and strength of the electric field at the same instant. Waves propagating along $+z$ $(-z)$ experience attenuation (amplification), consistent with Fig. 1a.

We present in Fig. 3b the detailed and exact dispersion [17, Eq. (S6)] of the edge states propagating along the $z$ direction. Without chiral-gain ($\varepsilon_{xz} = 0$), there is a spectral region with $\varepsilon_{\mathrm{m}}(\omega) < 0$ (shaded in gray) where surface plasmons cannot propagate. Curiously, in the presence of the chiral-gain response – no matter how small the gain parameter $\varepsilon_{xz}$ is – there is no longer a stopband for $\omega < \omega_{\mathrm{m}}$.

Figure 3c shows the projected dispersion in the complex $q$ plane for different chiral-gain values $\varepsilon_{xz}$. The curves are colored so that lighter (darker) tones correspond to lower (higher) frequencies. These results show that both plasmons propagating along the $+z$ ($q' > 0$) and $-z$ ($q' < 0$) directions are characterized by a positive $q''$. This implies that waves propagating along the $+z$ $(-z)$ direction are attenuated (amplified), consistent with Fig. 3a. Thus, the combination of the intrinsic spin-momentum locking of surface plasmons with the polarization-dependent non-Hermitian response of the chiral-gain material enables this system to act as a travelling wave amplifier.

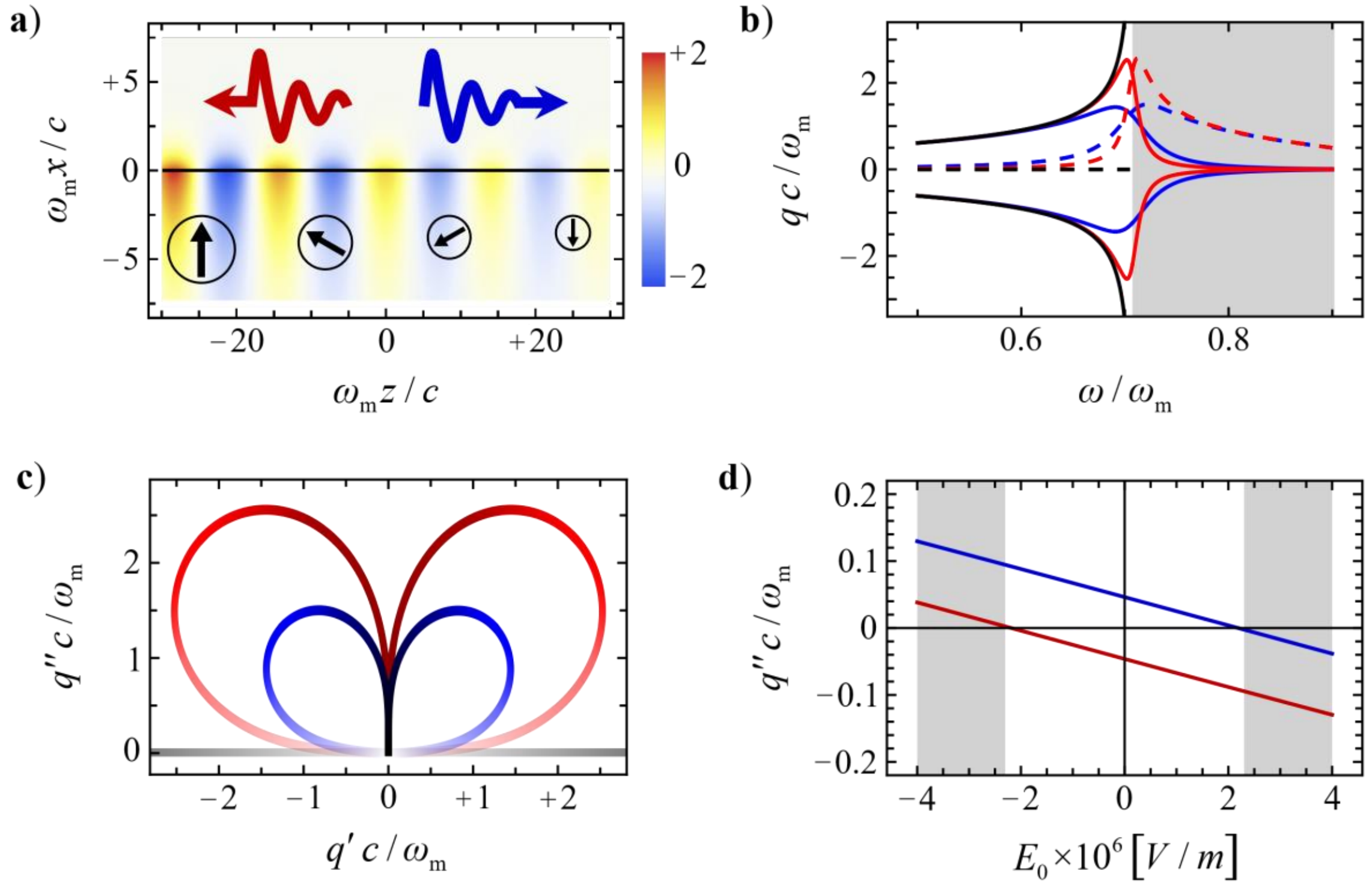

**Fig. 3 (a)** Snapshot in time ($t = 0$) of the magnetic field ($H_y$) associated with a surface plasmon propagating at the interface $x = 0$ between a dispersionless chiral-gain medium with $\varepsilon_{xx} = \varepsilon_{zz} = 1$, $\varepsilon_{xz} = 0.3$ ($x < 0$) and a lossless Drude metal ($x > 0$) for $\omega = 0.4\omega_{\mathrm{m}}$. The arrows provide a qualitative representation of the instantaneous electric field distribution. **(b)** Exact dispersion of surface plasmons propagating with $\omega \in \mathbb{R}$ at the interface represented in (a) for $\varepsilon_{xz} = 0$ (black), $\varepsilon_{xz} = 0.15$ (red) and $\varepsilon_{xz} = 0.3$ (blue). We represent the real/imaginary component of the wave number $q$ using solid/dashed curves. The imaginary part of $q$ is direction independent. The gray area represents a stopband for surface plasmons in the absence of chiral-gain ($\varepsilon_{xz} = 0$). **(c)** Projection of the exact dispersion of edge states in the complex wave number plane $q = q' + iq''$ for different values of the chiral-gain parameter: $\varepsilon_{xz} = 0$ (black), $\varepsilon_{xz} = 0.15$ (red), $\varepsilon_{xz} = 0.3$ (blue). Lighter/darker colors indicate smaller/larger frequencies in the range $0 < \omega < \omega_m$. **(d)** Imaginary part of the propagation constant $q$ of SPPs propagating along the $z$ direction over a flat interface $x = 0$ between an electrically biased slab of $WTe_2$ ($x < 0$) and a lossless Drude metal ($x > 0$) as a function of the static electric bias $E_0$. The plasma frequency of the Drude metal is $\omega_{\mathrm{m}} = 2\pi \times 100$ THz. The oscillation frequency is set at $\omega = 2\pi \times 45$ THz. The blue/red curve describes waves propagating along $+z/-z$, and the gray regions indicate the bias range where plasmons are amplified due to gain–momentum locking.

As previously mentioned, a promising route to engineer a chiral-gain medium relies on low-symmetry conductors with a Berry curvature dipole [6, 7]. In the supplementary information [17], we demonstrate that for a wide range of Weyl semimetals with the symmetry axis oriented along the *x*-axis, which include $MoTe_2$, and $WTe_2$ [9, 21], the electro-optic permittivity response is of the form $\varepsilon_{xx} = \varepsilon_{zz} = \varepsilon_{\infty} - \dfrac{\omega_{\mathrm{p}}^2}{\omega(\omega + i\gamma)}$ and $\varepsilon_{xz} \approx \dfrac{e^3}{\varepsilon_0 \hbar^2 \omega^2} D E_0$ where $D$ is the relevant component of the Berry curvature dipole tensor and the static bias is aligned with the *z*-direction. The formula assumes the high-frequency regime, $\omega >> \gamma$. Furthermore, we restrict our attention to the range of frequencies $\omega > \omega_{\mathrm{p}} / \sqrt{\varepsilon_{\infty}}$ to ensure operation above the plasma cutoff, i.e., the material has a dielectric-like response. As an example, we consider $WTe_2$ modeled by the following parameters: $D = -0.066$, $\omega_{\mathrm{p}} / \sqrt{\varepsilon_{\infty}} = 2\pi \times 12.09$ THz, $\gamma = 2\pi \times 7.25$ THz, and $\varepsilon_{\infty} = 3.3$ [21-23].

Figure 3d depicts the imaginary part of the propagation constant $q$ of the SPPs at the interface of an electrically biased slab of $WTe_2$ and a regular Drude metal as a function of the static electric bias $E_0$ at a fixed operating frequency. Consistent with the results in Fig. 2b, for a sufficiently strong bias the system exhibits gain-momentum locking and the direction of amplification is determined by the bias orientation. Since $D < 0$, waves propagating along $+z/-z$ are amplified if $E_0 > E_{th} / E_0 < -E_{th}$. Thus, Weyl semimetals such as $WTe_2$ provide an interesting option for realizing unidirectional systems with gain. They potentially offer several advantages over time-modulated systems, which may require high modulation speeds [24-29], magnetically biased systems that need bulky biasing circuits [30, 31], and systems employing drifting electrons, which demand large drift velocities [32-34].

To further illustrate the robustness of the surface plasmons with gain, we performed full-wave simulations using CST Studio Suite [35]. Figure 4 depicts a time snapshot of the plasmons at the interface of a metal and a chiral-gain medium for a circularly shaped edge. As expected, when waves are excited to propagate in the clockwise direction (Fig. 4a), the plasmons are attenuated by the chiral-gain material. Conversely, when waves are excited to propagate in the counter-clockwise direction, the plasmons are amplified (Fig. 4b).

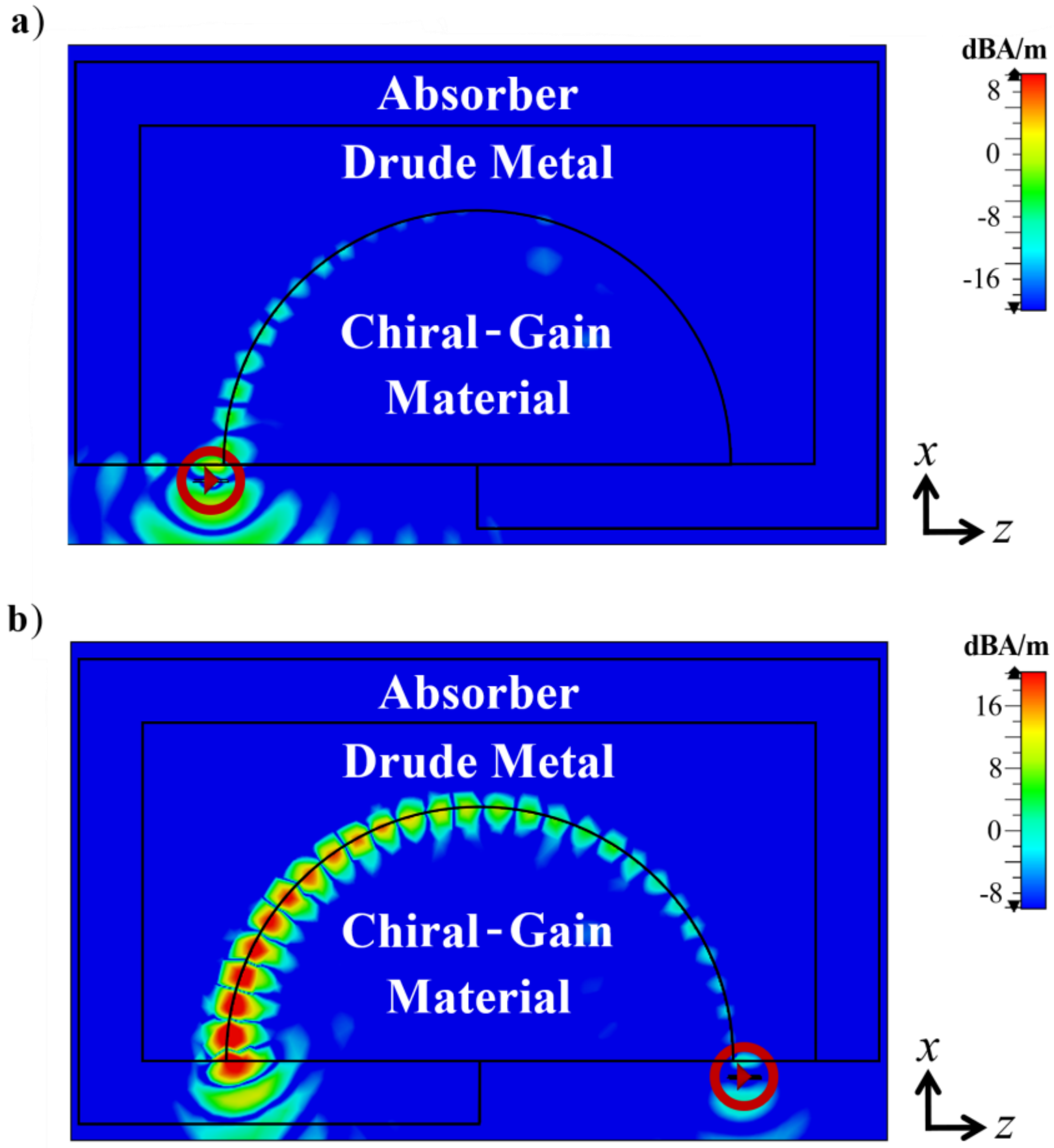

**Fig. 4** Full-wave simulations of the surface plasmons supported by the interface of a chiral-gain medium and a Drude metal. Excitation of (**a**) clockwise propagating plasmons and (**b**) counter-clockwise propagating plasmons. The chiral-gain medium is characterized by $\varepsilon_{xx} = \varepsilon_{zz} = 1$ and $\varepsilon_{xz} = 0.1$ at the operating frequency $\omega = 0.6\omega_{\rm m}$ with $\omega_{\rm m}$ the plasma frequency of the metal . We take $\Gamma = 0^{+}$ . The density plots represent a time snapshot of the magnetic field. The edge states are excited by half-wavelength dipoles located within the red circles.

In summary, we have characterized the propagation of edge states at the interface between a chiral-gain medium and a Drude metal using both quasi-static and exact formulations. We have shown that, despite the bulk response of the two materials being stable, the interface presents a gain-momentum locking effect. This phenomenon arises from the combination of the spin-momentum locking of surface plasmons and the polarization-dependent non-Hermitian response of the chiral-gain medium. We demonstrated how to leverage this property to design oscillators where the lasing modes have orbital angular momentum locked to the gain vector governing the chiral-gain, and to develop traveling wave amplifiers. Importantly, chiral-gain media based on electrically biased low-symmetry conductors such as $WTe_2$ can be used for loss-compensation of standard SPPs, enabling efficient propagation in photonic circuits. The gain only needs to be applied within the confined footprint of the SPPs, making this approach practical and feasible for real-world applications. For example, the guided SPP wavelength in Fig. 3d is $\lambda \sim 1\ \mu m$ at the gain threshold $E_{\text{th}} \sim 2.3\times 10^6\ V/m$. Thus, setting a bias field of this magnitude in a device with a width of 5 SPP wavelengths only requires a voltage of $11.5\ V$. Given the combination of nonreciprocity and gain within a single material, chiral-gain media hold significant promise for advancing nanophotonic circuitry.

## End Matter

*Time reversal symmetry*

The dispersionless chiral-gain medium exhibits time-reversal symmetry because $\bar{\varepsilon}(\omega) = \bar{\varepsilon}^*(\omega^*)$ [5, 8]. Therefore, in the case of a lossless Drude metal ($\Gamma = 0^+$), our edge waveguide is also time-reversal symmetric. This implies that the counter-propagating plasmons in Fig. 3a are connected through the time-reversal operation: if plasmons traveling along the $+z$ direction are characterized by a certain field distribution $(\mathbf{E}, \mathbf{H})$ and propagation constant $q$, then those traveling in the $-z$ direction are described by the reverse field distribution $\left(\mathbf{E}^*, -\mathbf{H}^*\right)$ with propagation constant $-q^*$ [36-38]. Consequently, the time snapshots at $t=0$ for counter-propagating plasmons appear qualitatively similar, meaning that the density plot in Fig. 3a accurately represents plasmons moving in both the $+z$ and $-z$ directions. Furthermore, $q''=\text{Im}\{q\}$ is independent of the propagation direction. We note that the spatial variation ($e^{+iqz}$) of the electric field in a time snapshot creates a perceived sense of rotation that is opposite to the actual time evolution of the electric field at a fixed point in space ($e^{-i\omega t}$). Waves propagating along the $+z$ direction exhibit left-elliptical polarization and experience attenuation, while those in the $-z$ direction have right-elliptical polarization and undergo amplification.

**Acknowledgements:** This work was partially funded by the Institution of Engineering and Technology (IET), by the Simons Foundation (Award SFI-MPS-EWP-00008530-10 for M.G.S. and Award SFI-MPS-EWP-00008530-04 to N.E.), and by FCT/MECI through national funds and, when applicable, co-funded EU funds under UID/50008: Instituto de Telecomunicações.

# Supplemental Information:

## *Gain-Momentum Locking in Chiral-Gain Media*

*João C. Serra[1,]*, Nader Engheta[2], Mário G. Silveirinha[1,]†*

*[(1)] University of Lisbon–Instituto Superior Técnico and Instituto de Telecomunicações, Avenida Rovisco Pais, 1, 1049-001 Lisboa, Portugal*
*[(2)] University of Pennsylvania, Department of Electrical and Systems Engineering, Philadelphia, PA 19104, U.S.A.*

The supplemental information provides additional details on A) bulk stability of the non-dispersive material, B) dispersion of the edge states at an interface between a chiral-gain medium and an isotropic material, C) quasi-static dispersion for the edge states propagating along a vertical interface, D) realization of a chiral-gain response relying on electrically biased low-symmetry conductors.

### *A. Bulk stability of the non-dispersive material*

In this supplementary note, we analyze the bulk response stability of the chiral-gain medium. For simplicity, we restrict our attention to the case of propagation in the $xoz$ plane. The bulk modes are plane waves $\mathbf{E}(\mathbf{r},t)=\mathbf{E}_0 e^{i\mathbf{k}\cdot\mathbf{r}}e^{-i\omega t}$. The corresponding dispersion equation is:

$$\mathbf{k}\times\left(\mathbf{k}\times\mathbf{E}_0\right)+\frac{\omega^2}{c^2}\overline{\varepsilon}\cdot\mathbf{E}_0=\mathbf{0}. \tag{S1}$$

For a relative permittivity tensor as in the main text

* E-mail: joao.serra@lx.it.pt
† To whom correspondence should be addressed: mario.silveirinha@tecnico.ulisboa.pt

$$\bar{\varepsilon} = \begin{pmatrix} \varepsilon_{xx} & 0 & \varepsilon_{xz} \\ 0 & \varepsilon_{yy} & 0 \\ 0 & 0 & \varepsilon_{zz} \end{pmatrix}, \tag{S2}$$

and for propagation in the $xoz$ plane, we find that the chiral-gain material supports two independent bulk modes:

$$\mathbf{E}_1 \sim e^{ik_x x} e^{ik_z z} e^{-i\omega t} \hat{\mathbf{y}}, \qquad \frac{\omega^2}{c^2} = \frac{k_x^2 + k_z^2}{\varepsilon_{yy}} \tag{S3a}$$

$$\mathbf{E}_2 \sim e^{ik_x x} e^{ik_z z} e^{-i\omega t} \left[ \left( k_x^2 - \varepsilon_{zz} \frac{\omega^2}{c^2} \right) \hat{\mathbf{x}} + k_x k_z \hat{\mathbf{z}} \right], \quad \frac{\omega^2}{c^2} = \frac{\varepsilon_{xx} k_x^2 + \varepsilon_{zz} k_z^2 + \varepsilon_{xz} k_x k_z}{\varepsilon_{xx} \varepsilon_{zz}}. \tag{S3b}$$

It is straightforward to check that for $\mathbf{k} = k_x \hat{\mathbf{x}} + k_z \hat{\mathbf{z}}$ real-valued, the eigenfrequencies are also real-valued provided $\varepsilon_{xx}\varepsilon_{zz} - \left( \frac{\varepsilon_{xz}}{2} \right)^2 > 0$, or equivalently $|\varepsilon_{xz}| < 2\sqrt{\varepsilon_{xx}\varepsilon_{zz}}$. It is relevant to note that for a stable system the eigenpolarizations of the material are linearly polarized. Hence, all the components of $\mathbf{E}_1$ ($\mathbf{E}_2$) are in-phase.

## *B. Dispersion of the edge states at an interface between a chiral-gain medium and an isotropic material*

Next, we deduce the exact edge dispersion for surface plasmons propagating at the flat interface (Fig. 1b of the main text) between a chiral-gain material and an isotropic material with relative permittivity $\varepsilon_{\mathrm{m}}(\omega)$.

For the horizontal interface $x = 0$ (with the chiral-gain material in the half-space $x < 0$), we look for transverse magnetic (TM) waves ($\mathbf{H} = H\hat{\mathbf{y}}$) with the following structure:

$$H\left(x,z,t\right)=H_0 e^{-i\omega t}e^{iqz}\times\begin{cases}e^{+\gamma_{\mathrm{C}}x}, & x<0\\ e^{-\gamma_{\mathrm{m}}x}, & x>0\end{cases}. \tag{S4}$$

Here, $\gamma_{\mathrm{m}}=\sqrt{q^2-\frac{\omega^2}{c^2}\varepsilon_{\mathrm{m}}\left(\omega\right)}$ is the attenuation constant (along $x$) in the isotropic region, whereas $\gamma_{\mathrm{C}}$ is the attenuation constant in the chiral-gain medium region. From Eq. (S3b), it is readily found that:

$$\gamma_{\mathrm{C}}=\sqrt{q^2\left(\frac{\varepsilon_{zz}}{\varepsilon_{xx}}-\frac{\varepsilon_{xz}^2}{4\varepsilon_{xx}^2}\right)-\frac{\omega^2}{c^2}\varepsilon_{zz}}-iq\frac{\varepsilon_{xz}}{2\varepsilon_{xx}}. \tag{S5}$$

The electric field can be obtained from Ampère's law as $\mathbf{E}=\frac{1}{-i\omega\varepsilon_0}\overline{\overline{\varepsilon}}^{-1}\cdot\left(\nabla H\times\hat{\mathbf{y}}\right)$. Therefore, the continuity of the tangential electric field ($E_z$) at the interface $x=0$ implies that $\frac{1}{\varepsilon_{zz}}\partial_x H$ is also continuous. This leads to the exact edge dispersion equation:

$$\frac{\gamma_{\mathrm{C}}}{\varepsilon_{zz}}+\frac{\gamma_{\mathrm{m}}}{\varepsilon_{\mathrm{m}}\left(\omega\right)}=0. \tag{S6}$$

Similarly, for the vertical interface $z=0$ (with the chiral-gain material in the half-space $z>0$), we look for TM waves characterized by the magnetic field distribution

$$H\left(x,z,t\right)=H_0 e^{-i\omega t}e^{iqx}\times\begin{cases}e^{+\gamma_{\mathrm{m}}z}, & z<0\\ e^{-\tilde{\gamma}_{\mathrm{C}}z}, & z>0\end{cases} \tag{S7}$$

with the attenuation constant $\tilde{\gamma}_{\mathrm{C}}=\sqrt{q^2\left(\frac{\varepsilon_{xx}}{\varepsilon_{zz}}-\frac{\varepsilon_{xz}^2}{4\varepsilon_{zz}^2}\right)-\frac{\omega^2}{c^2}\varepsilon_{xx}}+iq\frac{\varepsilon_{xz}}{2\varepsilon_{zz}}$. Imposing the continuity of the tangential electric field ($E_x$), we obtain the dispersion equation:

$$\frac{\tilde{\gamma}_{\mathrm{C}}}{\varepsilon_{xx}}+\frac{\gamma_{\mathrm{m}}}{\varepsilon_{\mathrm{m}}\left(\omega\right)}-iq\frac{\varepsilon_{xz}}{\varepsilon_{xx}\varepsilon_{zz}}=0. \tag{S8}$$

In the quasi-static limit ($|q| >> |\omega| / c$), one can use $\gamma_{\mathrm{C}} \approx \gamma_{\mathrm{C0}}$, $\tilde{\gamma}_{\mathrm{C}} \approx \tilde{\gamma}_{\mathrm{C0}}$ and $\gamma_{\mathrm{m}} \approx \gamma_{\mathrm{m0}}$ where $\gamma_{\mathrm{C0}}$, $\tilde{\gamma}_{\mathrm{C0}}$, $\gamma_{\mathrm{m0}}$ are defined as in the main text. Substituting these approximations into Eq. (S6) [Eq. (S8)] and solving with respect to $\omega$, it can be shown that the equation yields the same result as Eq. (4). Interestingly, this coincidence occurs even though the exact and quasi-static dispersion equations have quite different structures.

### *C. Quasi-static dispersion of the edge states for the vertical interface*

In this supplementary note, we characterize the dispersion of the edge states propagating along the (vertical) interfaces (perpendicular to the $z$ direction in Fig. 1b of the main text), using the quasi-static approximation. Since the system presents a two-fold rotation symmetry, we focus on the left-hand side interface ($z = 0$).

The quasi-static solution is constructed using the same steps as in the main text for the horizontal interface. The electric potential in the chiral-gain medium ($z > 0$) is $\phi(x,z,t) = \phi_0 e^{-i\omega t} e^{iqx} e^{-\tilde{\gamma}_{\mathrm{C0}} z}$, and in the metal region ($z < 0$) $\phi(x,z,t) = \phi_0 e^{-i\omega t} e^{iqx} e^{+\gamma_{\mathrm{m0}} z}$. Here, $q$ represents the propagation constant of the plasmons and $\gamma_{\mathrm{m0}}$ is defined as in the main text ($\gamma_{\mathrm{m0}} = \sqrt{q^2}$). The attenuation constant in the chiral-gain medium, $\tilde{\gamma}_{\mathrm{C0}} = \sqrt{q^2 \left( \frac{\varepsilon_{xx}}{\varepsilon_{zz}} - \frac{\varepsilon_{xz}^2}{4\varepsilon_{zz}^2} \right)} + iq \frac{\varepsilon_{xz}}{2\varepsilon_{zz}}$, is found by imposing $\nabla \cdot \left( \bar{\varepsilon} \cdot \nabla \phi \right) = 0$.

The quasi-static dispersion of the plasmons follows from the continuity of the normal component of the electric displacement vector, which yields:

$$\gamma_{\mathrm{m0}} \varepsilon_{\mathrm{m}}(\omega) + \tilde{\gamma}_{\mathrm{C0}} \varepsilon_{zz} = 0 . \qquad \text{(S9)}$$

Solving this equation with respect to the oscillation frequency for a real-valued $q$ in the $\Gamma = 0^{+}$ limit, one finds precisely the same result as in Eq. (4) of the main text.

### *D. Realization of a chiral-gain response relying on electrically biased low-symmetry conductors*

In this supplementary note, we discuss how chiral-gain responses like the ones considered in the main text can be implemented using electrically biased low-symmetry conductors.

We use the established link between the quantum geometry of a material and its transport properties under a static electric bias $\mathbf{E}_0$ [S1, S2]. Specifically, the velocity of Bloch electrons in low-symmetry crystals includes a contribution governed by the Berry curvature of the electronic bands. When a DC bias is applied, this anomalous velocity produces an electro-optic conductivity term given by

$$\bar{\sigma}_{\mathrm{EO}} = \frac{e^3}{\hbar^2 \gamma}\left[ -\left(\bar{\mathbf{D}}^T \cdot \mathbf{E}_0\right)\times \mathbf{1} + i\frac{\gamma}{\omega + i\gamma}\mathbf{E}_0 \times \bar{\mathbf{D}}^T \right], \tag{S10}$$

where $\bar{\mathbf{D}}$ denotes the (dimensionless) Berry curvature dipole tensor, $e$ the electron's charge, $\gamma$ the collision frequency and $\hbar$ is the reduced Planck's constant. Importantly, while the first term describes a conventional conservative gyrotropic response, the second term leads to non-Hermitian effects.

The full permittivity response consists of the conventional Drude-like term $\varepsilon_{\mathrm{D}}(\omega)=\varepsilon_{\infty}-\frac{\omega_{\mathrm{p}}^{2}}{\omega(\omega+i\gamma)}$, which for simplicity we assume to be isotropic, and the electro-optic susceptibility $\bar{\chi}_{\mathrm{EO}}(\omega)=\frac{\bar{\sigma}_{\mathrm{EO}}(\omega)}{-i\omega\varepsilon_{0}}$:

$$\bar{\varepsilon}(\omega)=\varepsilon_{\mathrm{D}}(\omega)\mathbf{1}+\bar{\chi}_{\mathrm{EO}}(\omega). \tag{S11}$$

We restrict our analysis to the high-frequency regime $\omega>>\gamma$, and suppose that $\omega>\omega_{\mathrm{p}}/\sqrt{\varepsilon_{\infty}}$ to ensure operation above the plasma cutoff. In addition, we ignore the first term in Eq. (S10) because it describes a conservative gyrotropic response, while our analysis is focused on the impact of the non-Hermitian (non-conservative) interactions associated with the second term. Under these approximations, Eq. (S11) reduces to:

$$\bar{\varepsilon}(\omega)\approx\mathbf{1}\left(\varepsilon_{\infty}-\frac{\omega_{\mathrm{p}}^{2}}{\omega(\omega+i\gamma)}\right)-\frac{e^{3}}{\varepsilon_{0}\hbar^{2}\omega^{2}}\left(\mathbf{E}_{0}\times\bar{\mathbf{D}}^{T}\right). \tag{S12}$$

To derive an electric response consistent with the nondispersive model of the MOSFET-metamaterial presented in the main text, it is convenient to consider materials with a rotation symmetry axis oriented along the $x$ direction. For such systems, the Berry curvature dipole tensor is constrained as follows [S3]:

$$\bar{\mathbf{D}}=\begin{pmatrix} D_{xx} & 0 & 0 \\ 0 & D_{yy} & D_{yz} \\ 0 & D_{zy} & D_{zz} \end{pmatrix}, \qquad \text{with } D_{xx}+D_{yy}+D_{zz}=0. \tag{S13}$$

In particular, for all the materials from the crystallographic point groups *mm*2, 3*m*, 4*mm* and 6*mm* the diagonal elements of the tensor vanish, i.e., $\bar{\mathbf{D}}=D_{yz}\hat{\mathbf{y}}\otimes\hat{\mathbf{z}}+D_{zy}\hat{\mathbf{z}}\otimes\hat{\mathbf{y}}$ [S3] (note that in Ref. [S3] the symmetry axis was taken the *z*-axis, while here it is the *x*-axis). Then, for an in-plane electric field aligned with the *z*-axis, $\mathbf{E}_0 = E_0\hat{\mathbf{z}}$,

$$\mathbf{E}_0 \times \overline{\mathbf{D}}^T = -DE_0\hat{\mathbf{x}} \otimes \hat{\mathbf{z}}$$

with $D = D_{zy}$ (we assume that the in-plane axes $y$ and $z$ are chosen such $|D_{zy}| > |D_{yz}|$ to maximize the linear electro-optic effect). The corresponding reduced permittivity tensor (S12) is given by

$$\bar{\varepsilon}(\omega) = \begin{pmatrix} \varepsilon_{\mathrm{D}} & 0 & \varepsilon_{xz} \\ 0 & \varepsilon_{\mathrm{D}} & 0 \\ 0 & 0 & \varepsilon_{\mathrm{D}} \end{pmatrix}, \tag{S14a}$$

exactly as in the non-dispersive model in the main text, but now the permittivity elements are frequency dependent:

$$\varepsilon_{\mathrm{D}} = \varepsilon_\infty - \frac{\omega_{\mathrm{p}}^2}{\omega(\omega + i\gamma)} \quad \text{and} \quad \varepsilon_{xz} = \frac{e^3}{\varepsilon_0 \hbar^2 \omega^2} DE_0 . \tag{S14b}$$

Note that, similar to the MOSFET-metamaterial [S4], $\varepsilon_{xz}$ is real-valued.

In summary, for the materials from the crystallographic point groups *mm*2, 3*m*, 4*mm* and 6*mm* the linear electro-optic response is described by the above formula when $\omega >> \gamma$. Examples of such materials include several Weyl semimetals such as TaAs, TaP, NbAs, NbP, $MoTe_2$, and $WTe_2$, for which the Berry dipole $D$ can range from 0.03 (TaP) up to 20 (NbP) [S5]. Moreover, materials from the $\bar{4}2m$ point group with a rotation symmetry axis oriented along the $x$ direction subject to an electric bias $\mathbf{E} = E_0\hat{\mathbf{y}}$ present a similar response.

To illustrate how the electro-optic response affects the propagation of SPPs, next we consider a flat interface between an electrically biased slab of WTe2 ($x<0$) and a lossless Drude metal ($x>0$) characterized by the permittivity $\varepsilon_{\rm m}(\omega)=1-\frac{\omega_{\rm m}^2}{\omega^2}$ as in Fig. 1a of the main text. The edge state dispersion is given by Eq. (S6), with the permittivity tensor of the chiral-gain material defined as in Eq. (S14). The material response of WTe2 is modeled with parameters obtained from Refs. [S5-S7]: $D=-0.066$, $\omega_{\rm p}/\sqrt{\varepsilon_\infty}=2\pi\times12.09$ THz, $\gamma=2\pi\times7.25$ THz (at room temperature), and $\varepsilon_\infty=3.3$.

Figure S1 (solid line) depicts the imaginary part of the propagation constant $q$ as a function of the static electric bias $E_0$. Consistent with the results for the nondispersive model in the main text, for a sufficiently strong bias, the systems exhibits gain-momentum locking and the direction of amplification is determined by the bias orientation, i.e., waves propagating along $+z$ ($-z$) are amplified when $E_0>E_{\rm th}$ ($E_0<-E_{\rm th}$). The threshold bias $E_{\rm th}>0$ depends on the frequency of oscillation $\omega$.

For completeness, we compare the results obtained with the complete permittivity tensor (S11) (dashed curves in Fig. S1) with the reduced response (S14) (solid curves in Fig. S1). As seen, the simplified model correctly predicts the overall behaviour of the dispersion relation and, in particular, the value of threshold bias to achieve plasmon amplification.

We would like to underline that the electric bias only needs to be applied along the footprint of the surface plasmons, across a distance comparable with the SPP's wavelength. For example, the guided SPP wavelength in Fig. S1 is $\lambda\sim1\ \mu m$ at the gain threshold $E_{\text{th}} \sim 2.3\times10^{6}\ V/m$. Thus, setting a bias field of this magnitude in a device with a width of 5 SPP wavelengths requires a voltage of $11.5\ V$.

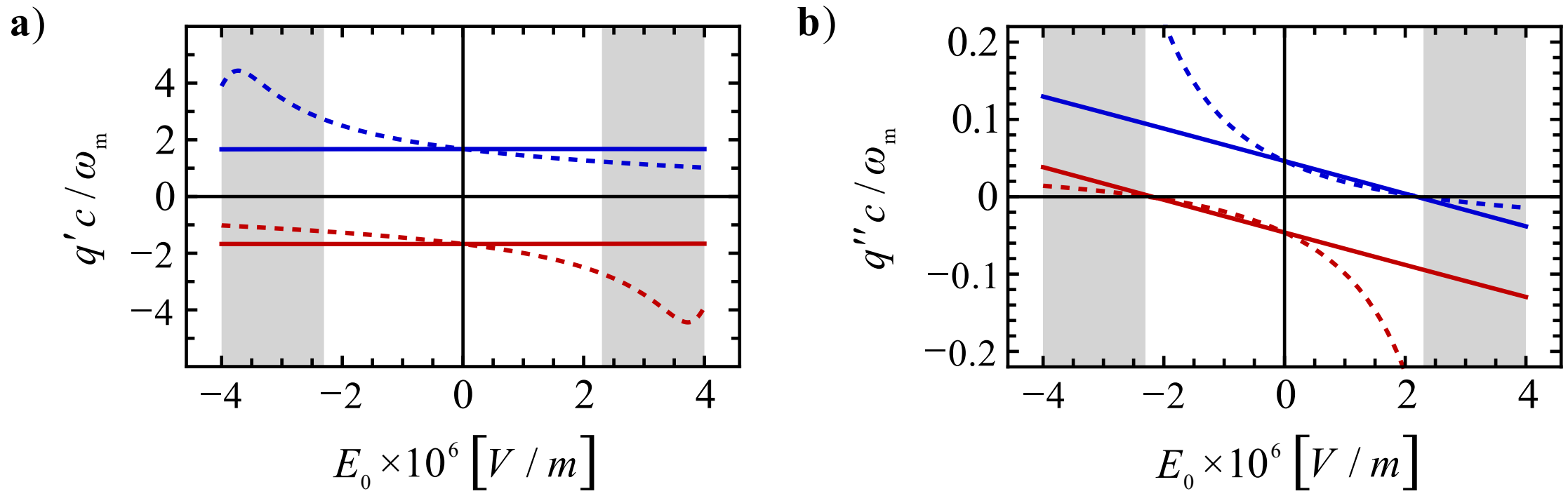

**Fig. S1** Real (a) and imaginary (b) part of the propagation constant $q$ of SPPs moving along the $z$ direction over a flat interface $x=0$ between an electrically biased slab of $WTe_2$ ($x<0$) and a lossless Drude metal ($x>0$) as a function of the static electric bias $E_0$. The $WTe_2$ slab is characterized by the parameters $\varepsilon_\infty = 3.3$, $\omega_{\text{p}} = 2\pi\times21.96$ THz, $\gamma = 2\pi\times7.25$ THz, $D = -0.066$. The Drude metal has a plasma frequency $\omega_{\text{m}} = 2\pi\times100$ THz. The oscillation frequency is set at $\omega = 2\pi\times45$ THz. The blue/red curve describes waves propagating along $+z/-z$ for the full (S11) (dashed line) and reduced (S14) (solid line) permittivity responses. The grey regions indicate the bias range where plasmons are amplified due to gain–momentum locking.

## Supplemental Information References